# Learning to Car-Follow Using an Inertia-Oriented Driving Technique: A Before-and-After Study on a Closed Circuit

Kostantinos Mattas[1], Antonio Lucas-Alba[2], Tomer Toledo[3], Óscar M. Melchor[4], Shlomo Bekhor[3], Biagio Ciuffo[1],

[1]*European Commission, Joint Research Centre, Ispra, VA, Italy*

[2]*Departament of Psychology and Sociology, Universidad de Zaragoza, C/Ciudad Escolar s/n, 44003, Teruel, Spain (lucalba@unizar.es)*

[3]*Department of Civil and Environmental Engineering, Technion – Israel Institute of Technology, Haifa, Israel*

[4] *Impactware, Madrid, Spain*

**Abstract**

**Background**. For decades, car-following (CF) and traffic flow models have assumed that drivers' default driving strategy is to maintain a safe distance. Several previous studies have questioned whether the Driving to Keep Distance (DD) is a traffic invariant; therefore, the acceleration–deceleration torque asymmetry of drivers must necessarily determine the observed patterns of traffic oscillations. Those studies indicate that drivers can adopt alternative CF strategies, such as Driving to Keep Inertia (DI), by following basic instructions. The present work extends the evidence from previous research by showing the effectiveness of a DI course that immediately translates into practice on a closed circuit. **Methods**. Twelve drivers were invited to follow a lead car that varied its speed on a real circuit; then, the driver took a DI course and returned to the same real car-following scenario. **The results**. Drivers generally adopted DD as the default CF mode in the pretest, both in field and simulated PC conditions, yielding very similar results. After taking the full DI course, drivers showed significantly less acceleration, deceleration, and speed variability than did the pretest, both in the field and in the simulated conditions, which indicates that drivers adopted the DI strategy. This study is the first to show the potential of adopting a DI strategy in a real circuit.

1. **Introduction**

Most car-following (CF) models assume that Driving to Keep Distance (DD) is a traffic invariance, that is, a predictable and recurring factor that is part of the overall behavior of drivers in a platoon (Pariota et al., 2016; Sharma et al., 2019; Toledo, 2007; Saifuzzaman and Zheng, 2014). However, as shown by recent research, both humans (Blanch et al., 2018; Taniguchi et al., 2015) and automated vehicles (Stern et al., 2019; Stern et al., 2018) can adopt alternative car-following strategies to improve traffic flow. Focusing on the human driver, it is important to highlight that DD is not consubstantial or endogenous to the driver; it is a taught and learned CF strategy. However, drivers can learn to avoid sudden variations in speed by not driving too close to a lead car that varies speed and maintaining a speed with minimal oscillations, i.e., Driving to Keep Inertia (DI).

A series of studies have shown that drivers adopt the DD technique by default but can easily adopt the DI technique following short and simple direct instructions (Blanch et al., 2018; Lucas-Alba et al., 2020). A more ambitious objective is for drivers to understand how traffic jams form and learn for themselves how to prevent them from spreading; with that objective, the WaveDriving Course (WDC, Melchor et al., 2018; Lucas-Alba et al., 2022) was designed. The WDC has recently been evaluated using a methodological design that includes a control group and an independent driving simulator into which the WDC is integrated, yielding very positive results (Tenenboim et al., 2022). When DI and DD are compared under heavy traffic, the DI strategy becomes more efficient, resulting in less dispersion of speed and lower fuel consumption (Blanch et al., 2028; Lucas-Alba et al., 2020; Tenenboim et al., 2022). If the concept of eco-driving implies less acceleration and deceleration while maintaining stable speed (Boriboonsomsin et al., 2010; Haas and Bekhor, 2017), the DI strategy is a tool for understanding why and how to drive in an ecological way to avoid causing traffic jams.

This study aimed to test whether drivers without prior knowledge of the DI strategy can learn about the DI strategy and then apply it by driving a real car in a controlled driving scenario (closed circuit). Previous studies have described how DD and DI differ and how DI benefit from DI driving, confirming that drivers can adopt DD or DI to follow a leader who fluctuates in speed and that these techniques show opposite patterns of speed and distance: DD present high speed variability and low distance variability, while DI present low speed variability and high distance variability (Blanch et al., 2018; Lucas-Alba et al., 2020). More recent studies have shown that drivers can drive in a driving simulator, learn about DI technique by following the WDC, and then drive again in a driving simulator and that this learning is productive (Tenenboim et al., 2022). The present study represents an extension of this work, using the simulated environment of the WDC to evaluate DI learning but combining this learning and the corresponding evaluation in the simulator with a measurement in real vehicles obtained before and after the course.

## 2. Methodology

The study is based on a test campaign carried out on two consecutive days on the road infrastructure of the Joint Research Centre (JRC) in Ispra (Italy), the third largest European Commission site. Participants followed a pretest/intervention/posttest design. Participants were divided into three teams. Each team formed part of a platoon of six vehicles. The first vehicle was driven by one of the researchers in the study who followed a repetitive pattern of acceleration and deceleration between 30 and 45 km/h, simulating traffic congestion. The second vehicle was also driven by a researcher, and its subsequent speed was autonomously regulated by an adaptive cruise control (ACC) system with the intent of amplifying the stop-and-go mode. Vehicles 3 to 6 were driven by the experiment participants. The participants were randomly assigned to each position and unaware of the study objectives. Drivers were told not to overtake, and their vehicles did not have any driving assistance. After a contact lap with the assigned vehicle and the route, the participants made two complete laps of a pre-established route. The drivers were then taken to a classroom where they received the WDC (delivered in English), which lasted approximately 50–60 minutes. The course included an instructional session with a researcher and carried out practice using a laptop computer. Finally, the drivers returned to the vehicles and performed the same CF exercise as that described above, along the same route and in the same order.

### 2.1. Goals

The study aimed to determine whether A) drivers adopt a DD strategy in the pretest as a default, both in the real setting and when using the WDC simulator, and B) drivers change their driving patterns after full experience with the DI course (tutorials + practice) in the posttest, both when using the WDC simulator and when using real cars, i.e., adopting a DI strategy in the posttest in terms of speed, distance and fuel consumption.

## 2.2. Participants

A total of 12 driver participants (eight men and four women) were recruited by invitation at the JRC premises. Participants included postdoctoral students, JRC staff, and acquaintances, all of whom held at least a university degree and represented various nationalities (Colombian, Greek, Hungarian, Italian, and Spanish). All 12 drivers had normal or corrected-to-normal vision. The ages of the participants ranged from 29 to 43 years, with an average of 33.6 years (SD = 4.86). Regarding driving experience, the participants' possession of driver licenses varied between 9 and 25 years, with an average of 15.0 years (SD = 5.85). Five of the participants reported driving less than 10000 km/year, and another five reported driving between 10000 and 20000 km/year. Five reported that they drive both on urban and interurban roads, and the other five reported that they mostly drive on interurban (highway) roads. All participants were well acquainted with typical driving conditions, and none of them were novice drivers.

## 2.3. Design

The present study included a pretest/intervention/posttest design. The participants were divided into three teams with four driver participants each who operated in three consecutive independent sessions. In each session, two researchers drove the first two cars as described above. All participants followed the same procedure (drive a real car/evaluate in the WDC module/take specific WDC lessons/evaluate in the WDC module/drive a real car). The study received approval from the JRC ethics committee (number 32904-1-26102023).

## 2.4. Materials

This study required two sets of complementary materials: six instrumented vehicles for the on-road experiment and laptops and WDC software in the cloud for the training session.

### 2.4.1 Driving with real cars in a closed circuit

**Figure 1** shows the circuit in which the six-vehicle convoy was driving. Its total length was 3.3 km. An average of 7.3 minutes was needed to complete one lap. Each vehicle was equipped with several sensors that were able to track its exact position and speed. As indicated in Figure 1, there were several sharp curves in the circuit in which drivers had naturally decreased their speed.

The experiments were conducted on May 31 (one team) and June 1 (two teams) 2024. These were sunny and dry days. In the pretest, drivers were asked to follow the car in front of them without overtaking as they would normally do. After the pretest, each team of drivers undertook the WDC, which was administered in a classroom close to the circuit. After completing the course, the participants returned to the same circuit (i.e., the posttest drive). The drivers received the same driving instructions to follow the vehicle in front without overtaking it, as in the pretest. Specifically, they were not instructed as to which driving technique to employ. The duration of the entire experiment (pretest, course, and posttest) was approximately 2 hours for each team.

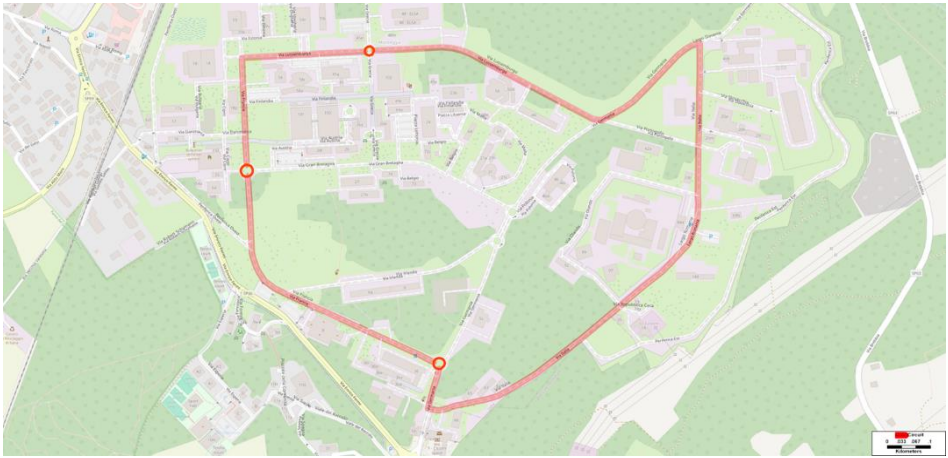

**Fig. 1.** The circuit followed at the JRC facilities in Ispra (the circles represent roundabouts).

**2.4.2 Learning an inertia-oriented driving technique: DI course**

The DI course has two main purposes. The first is to teach an alternative CF technique, which focuses not only on maintaining safety distance (DD) but also on preserving inertia when following an oscillating leader (DI). Drivers are typically taught to drive to maintain distance (DD), which enlarges disturbances along the platoon. The alternative to coping with a lead oscillatory vehicle (the shockwave origin) is the Driving to Keep Inertia (DI) strategy, i.e., anticipating the stop-and-go pattern and becoming shockwave proofing offsetting or damping waves and maintaining a uniform speed (Blanch et al., 2018; Lucas-Alba et al., 2020). Accordingly, drivers taking the DI course are encouraged to compare the differing effects of adopting DD versus DI in different tasks and scenarios. The second main purpose of the DI course is to help drivers understand the close connection between adopting DD or DI and the emergence of congestion in a platoon of vehicles. To achieve these goals, several ordinary elements and circumstances (e.g., a rearview mirror, the presence of traffic lights, or passing road sections with different speed limits) have been implemented in the DI simulator, as well as some very unusual or impossible elements and circumstances (e.g., adopting bird's-eye views from different angles and positions across the whole platoon at will, displaying traffic lights on every single car, connecting cars with springs, and activating radar-like displays; see Lucas-Alba et al., 2022). The WDC consists of 6 modules, each of which begins with a video tutorial (explaining the specific objectives of the module) followed by a practice period in the simulator. The WDC follows this sequence (**Figure 2**): Knowing the controls (getting to know the simulator: Module 0); Pretest Evaluation (Module E1); Teaching DI (Modules 1-3); Posttest Evaluation (Module E2). Pre- and posttest evaluations were performed via the same procedure. The sessions in the course are time-limited (between 2.5 and 5 min), and video tutorials take no longer than this time range to address DI principles and what the learner should practice at each level. Thus, the time a learner would need to complete the DI course (tutorials + practice on the WDC simulator) is approximately 50-60 min (Lucas-Alba et al., 2022; Tenenboim et al., 2022).

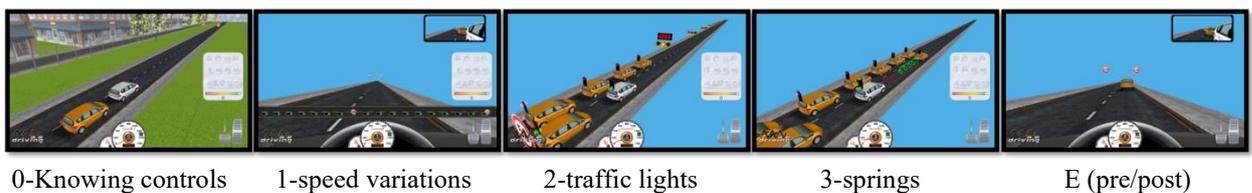

 0-Knowing controls    1-speed variations    2-traffic lights    3-springs    E (pre/post)
**Fig. 2.** The WDC learning modules

### 2.5. Procedure

All participants began the experiment by providing informed consent and completing the detailed questionnaire. Participants were uninformed about the purpose of the study. The participants were told that the objective of the study was to examine different responses in a (simulated) driving scenario. The experimenter then briefly described the different experimental stages.

In the first stage, the experimenter brought participants to the parking lot where the cars were available. All cars were instrumented with GNSS positioning instrumentation and laptops to store individual speed and position. The participants were randomly assigned to positions. They were simply told that they had to follow the car right in front of them. Participants performed a test ride and then completed two laps of the predetermined course shown in **Figure 1**. This stage took approximately 30 minutes.

In the second stage, participants were taken to a teaching room with a large rectangular table, chairs and laptops. The participants were then introduced to the experimenter in charge of directing the WaveDriving Course session. The participants were given passwords to access the WDC software online using laptops. The WDC has six modules. Each module began with a tutorial shown on a large screen and explained by the experimenter to guide and facilitate practice on the simulator. These tutorials play two fundamental roles. On the one hand, they specify the objectives of each module. On the other hand, they model the activity in the simulator (i.e., they propose which courses of action should or should not be followed, providing examples; Bandura, 1986). Modeling is particularly important in Modules 1–3, where DD-type activity (left lane) is compared with DI-type activity (right lane). After each tutorial, the practice concerning that module took place. The Teaching DI modules are especially important: Module 1 focuses on demonstrating how cyclical variations in speed (e.g., between 60 and 20 km/h) affect the stability of the following platoon. Module 2 highlights the benefits of identifying a uniform speed that allows drivers to pass through a series of traffic light cycles without having to stop. Module 3 introduces the concept of an anti-jam distance—a variable distance that enables the maintenance of a uniform speed even when following a leader who alternates between acceleration and braking cycles. The evaluation is conducted both before and after Modules 1 to 3 and involves following a lead vehicle that exhibits a cyclical pattern of acceleration and deceleration, typical of peak-hour traffic and congestion. A detailed description of the modules can be found in Lucas-Alba et al. (2022).

After the evaluation, each participant received their electro-car-diagram (ECD), a play on words inspired by ECG that was designed to convey that this tool is a personalized diagnostic tool for drivers. These diagrams were provided both before and after the Teaching DI modules. The ECDs (see examples in Figures 3 and 4) present the participants' driving activity across six individual graphs, each summarizing performance on a key variable. These variables, displayed from left to right and top to bottom, are as follows: speed, fuel consumption, distance traveled, personal safety relative to chosen speeds, personal safety based on distance to the lead vehicle, and collective safety. A group of eight virtual DD followers is considered.

**Figure 3** presents an example of a participant's results prior to the Teaching DI modules. In this trial, the participant followed a lead vehicle exhibiting a stop-and-go pattern. The participants' driving behavior reflects this, as shown in the SPEED (top-left) and FUEL (top-middle) consumption graphs, which display pronounced peaks and valleys—a clear indication of a DD strategy. The DISTANCE graph (top-right) shows the total distance traveled over time. When the distance increases as the test progresses, it follows a stepped pattern, with flat segments indicating frequent complete stops, which reflects poor space effectiveness. In the SPACE-SAFETY graph (bottom-left), two lines are shown. The green dashed line represents the distance that should be kept based on the participant's current speed. The yellow line shows the actual distance maintained. This line often deviates significantly—either too long or too short—suggesting inconsistent safety margins. The SAFETY FRONT graph (bottom-middle) depicts the participant's distance to the vehicle ahead. The flat valleys in this graph indicate that the distance fell below the safe threshold, indicating possible near-collision scenarios. Finally, the SAFETY REAR graph (bottom-right) displays the extension of the platoon composed of eight virtual DD followers behind the participant. Fluctuations—peaks and valleys—reflect dynamic changes in the spacing among the following vehicles. This "accordion-like" behavior implies that reductions in inter-vehicular distance (from peak to valley) are accompanied by sudden decelerations, which can potentially lead to rear-end collisions.

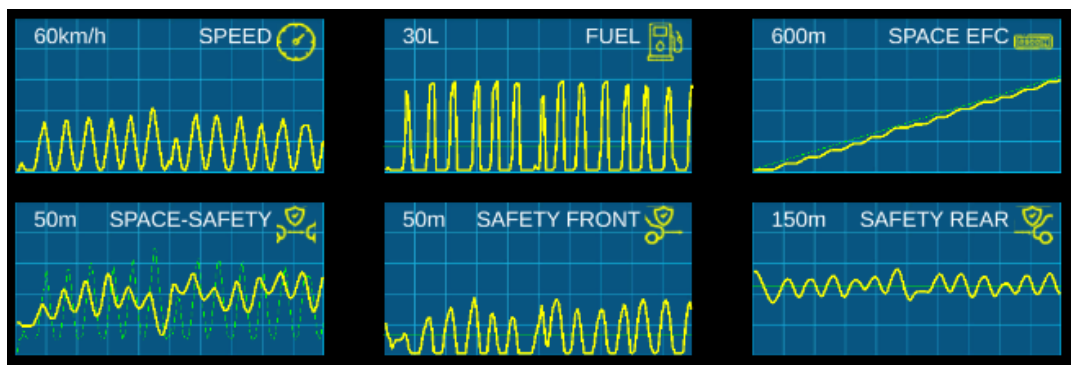

**Fig. 3.** Example of an ECD by a participant before DI learning (Modules 1-3)

**Figure 4** presents an example of a participant's performance after completing the Teaching DI modules. As in the previous example, this participant drove behind a leader exhibiting a stop-and-go pattern. However, the SPEED (top-left) and FUEL consumption (top-middle) graphs reveal a markedly different strategy: the participant maintained steady car-following behavior characterized by minimal speed variation and reduced fuel consumption. The DISTANCE graph (top-right) shows a smooth and continuous increase in distance traveled, indicating uninterrupted movement throughout the trial. In the SPACE-SAFETY graph (bottom-left), the participant successfully maintained appropriate spacing: there were no apparent conflicts between the actual distance and the ideal distance based on speed. This is further confirmed by the SAFETY FRONT graph (bottom-middle), which shows no flat valleys— indicating that the driver consistently maintained a safe following distance, with no moments of elevated collision risk. Finally, the SAFETY REAR graph (bottom-right) reveals a very uniform inter-vehicular spacing among the eight virtual DD followers—unlike the accordion-like pattern described earlier. This consistency suggests a stable platoon dynamic, with minimal fluctuations in speed and spacing. As this example illustrates, the participant adopted a DI-based strategy that is fundamentally different from the DD pattern shown in Figure 3. Two distinct car-following strategies can emerge in response to the same leader behavior (see Lucas-Alba et al., 2025).

After completing this stage, the participants returned to the parking lot, took the car they had previously used and the same position on the platoon to complete two laps of the same circuit. Finally, the participants were debriefed. The duration of the entire experiment was approximately 120 minutes for each team of participants.

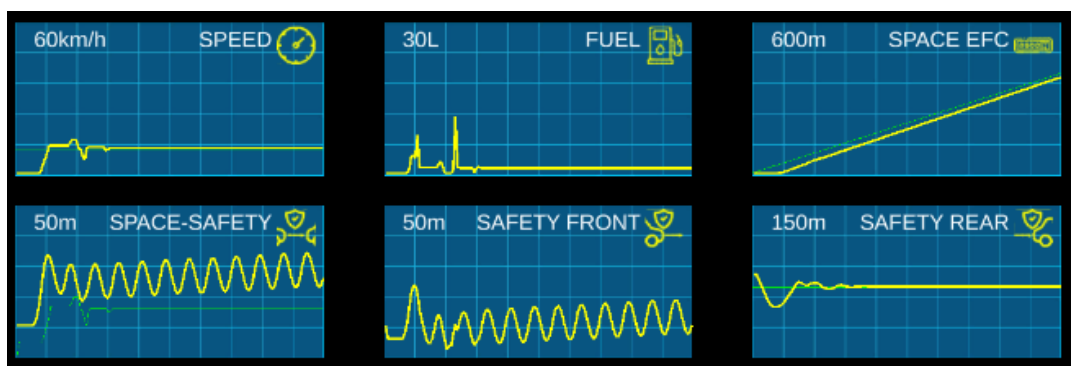

**Fig. 4.** Example of an ECD by a participant after DI learning (Modules 1-3)

### 3. Results

Below, we present the results of the effects of the DI course on the driving behavior of the participants, both in the WDC simulator itself and with real cars on the circuit at the JRC.

#### 3.1 Driving simulator

During the reception of the WDC, many problems were observed with the laptop of one of the participants. Therefore, only the WDC data of 11 participants were analyzed (**Table 1**). The WDC evaluation module invites participants to follow a lead vehicle as they see fit (Lucas-Alba et al., 2022; Tenenboim et al., 2022). That leader vehicle followed a trajectory that we can consider characteristic of dense and congested traffic, with rather cyclical acceleration and deceleration patterns. Eight virtual vehicles circulated behind the participants, and all of them followed a DD strategy. The analysis in Table 1 compares the driving performance between the pretest and the posttest, focusing on four variables: speed, distance to the leading vehicle, extension of the platoon of following vehicles and fuel consumption. These results largely coincide with those obtained in other studies. No differences were observed in the mean speed between the pretest and posttest, but there were differences in dispersion, which was lower in the posttest. This lower speed variability in the posttest also translates into lower fuel consumption. A lower variability in speed during the posttest does not generate changes in the average extent of the platoon of followers, but it does translate into less fluctuation in the platoon length (dispersion) along the route during the posttest. Finally, the average following distance to the leading vehicle does not vary between the pretest and the posttest, but its dispersion does, which is lower in the posttest. This is an unexpected result that we address in the discussion section.

**Table 1.** Results of the pretest and posttest evaluations of the DI learning simulator.

| Variable | Pretest (M, SD) | Posttest (M, SD) | t-student (gl, $p$) | Cohen d |
|---|---|---|---|---|
| Speed: average (m/s) | 10.01 (2.93) | 10.20 (2.49) | -.995 (10, .343) | -.30 |
| Speed: standard deviation (m/s) | 7.59 (1.78) | 3.20 (2.35) | 4.83 (10, .001) | 1.40 |
| Distance: average (m) | 9.34 (7.85) | 11.30 (6.56) | -.883 (10, .398) | -.27 |
| Distance: standard deviation (m) | 11.01 (3.02) | 6.94 (2.22) | 4.08 (10, .002) | 1.23 |
| Fuel consumption (l) | 569.78 (269.74) | 224.17 (161.27) | 3.69 (10, .004) | 1.11 |
| Platoon: average (m) | 69.39 (8.40) | 69.85 (7.24) | -.541 (10, .601) | -.16 |
| Platoon: standard deviation (m) | 7.75 (1.61) | 3.68 (2.25) | 4.33 (10, .001) | 1.31 |

### 3.2 Real cars on a closed circuit

As indicated in the methodology section, the first two vehicles in the platoon were driven by the research team. We are interested in the behavior of the last 4 cars in the platoon (Vehicles 3-6) for each team. Due to technical complications (specifically, the loss of records), the overall sample presents additional challenges for data analysis. For the first team, only the results for drivers 3 and 4 were available; for the second team, the results were obtained for drivers 3, 4 and 6; and for the third team, the data were available for all four drivers 3 through 6 (**Table 2**). Furthermore, the first team completed two full laps of the circuit, whereas the second and third teams completed three laps. This distribution of records was consistent across both the pretest and posttest phases.

**Table 2.** Summary of the data collected in the JRC circuit loop experiment.

| Team | Session | Total Laps | Vehicles with data collected | Total observations | Observations with WDC drivers |
|---|---|---|---|---|---|
| 1 | Pretest | 2 | 1,3,4 | 6 | 4 |
| 1 | Posttest | 2 | 1,3,4 | 6 | 4 |
| 2 | Pretest | 3 | 1,3,4,6 | 12 | 9 |
| 2 | Posttest | 3 | 1,3,4,6 | 12 | 9 |
| 3 | Pretest | 3 | 1,2,3,4,5,6 | 18 | 12 |
| 3 | Posttest | 3 | 1,2,3,4,5,6 | 18 | 12 |

**Table 3** presents the first evaluation of the results after Student's t test was used to compare the pretest and posttest scores. These results significantly overlap with the data obtained before and after the WDC (e.g., fuel consumption, speed dispersion, distance to the leader). However, compared to the simulator data (Table 1), these data present more complex characteristics: first, the scores are not fully independent (e.g., Driver 4's performance may be influenced to some extent by Driver 3's performance); second, the measurements are repeated across the pretest and posttest; third, some data are missing; and, finally, the records are grouped into three independent blocks (corresponding to the three teams).

**Table 3.** Results of the pretest and posttest evaluations at the closed circuit (*Student's t test*).

| Variable | Pretest (M, SD) | Posttest (M, SD) | t-student (gl, *p*) | Cohen d |
|---|---|---|---|---|
| Speed: average (m/s) | 7.70 (.79) | 7.79 (.42) | -.622 (24, .27) | -.12 |
| Speed: standard deviation (m/s) | 2.41 (.78) | 2.09 (.50) | 1.71 (24, .05) | .34 |
| Distance: average (m) | 20.35 (5.04) | 26.09 (7.10) | -3.10 (24, .002) | -.62 |
| Distance: standard deviation (m) | 7.52 (2.71) | 9.98 (3.71) | -3.33 (24, .001) | -.67 |
| Fuel consumption (l) | 10.06 (2.25) | 8.18 (2.24) | 6.30 (24, .001) | 1.26 |
| Platoon: average (m) | 59.97 (10.31) | 78.87 (17.84) | -1.75 (5, .071) | -.71 |
| Platoon: standard deviation (m) | 16.35 (3.14) | 21.35 (5.80) | -3.75 (5, .007) | -1.53 |

To address these limitations in evaluating changes in driving performance before and after the training course, a linear mixed-effects model (LMM) was fitted using SPSS version 29. The dependent variables were average speed, dispersion of speed, average distance to leader, dispersion of distance to leader, average platoon length and dispersion, and fuel consumption. The fixed effects included time (pretest vs posttest), team (three independent driving groups), and lap (number of laps completed). The driver was included as a random effect to account for within-subject repeated measures. Repeated observations were specified across combinations of Time and Lap, with a compound symmetry (CS) covariance structure. The model was estimated using restricted maximum likelihood (REML) and Satterthwaite approximations for degrees of freedom. The estimated marginal means (EMMEANS) were computed for the time factor, with pairwise comparisons adjusted using the least significant difference (LSD) method.

A preliminary model including the Time × Team × Lap interaction revealed the statistical significance of average speed, speed dispersion, average distance and fuel consumption (marginal significance for distance dispersion); however, this interaction was excluded from the final model due to the absence of an a priori hypothesis, the limited sample size, and the potential risk of overfitting. The final model focused on the main effect of Time (Pretest vs Posttest) while controlling for Team and Lap, as these factors were included to increase data density rather than to test theoretical contrasts. No main effects of Team or Lap were of theoretical interest and were retained in the model solely as control variables. **Table 4** presents the main statistics obtained with the LMM procedure.

**Table 4.** Results of the pretest and posttest evaluation at the closed circuit (*LMM*).

| Variable | Pretest (M, SE) | Posttest (M, SE) | F (gl, *p*) | 95% CI |
|---|---|---|---|---|
| Speed: average (m/s) | 7.71 (.075) | 7.80 (.075) | .48 (1, 38.97, .49) | -.35, .17 |
| Speed: standard deviation (m/s) | 2.49 (.93) | 2.17 (.93) | 4.15 (1, 38.97, <.05) | .002, .643 |
| Distance: average (m) | 21.21 (1.47) | 26.91 (1.47) | 12.99 (1, 38.07, <.001) | -8.90, -2.50 |
| Distance: standard deviation (m) | 8.07 (0.78) | 10.53 (0.78) | 12.23 (1, 38.20, .001) | -3.89, -1.04 |
| Fuel consumption (l) | 9.62 (0.67) | 7.74 (0.67) | 46.93 (1, 38.03, <.001) | 1.33, 2.44 |
| Platoon length: average (m) | 59.97 (5.95) | 78.87 (5.95) | 5.05 (1, 10, <.05) | -37.65, -.154 |
| Platoon length: dispersion (m) | 16.35 (3.74) | 21.35 (3.74) | 12.00 (1, 10, .007) | -8.27, -1.74 |

The results in Table 4 follow the expected trends. The analysis of average speed did not reveal a significant main effect of time. Changes were observed for speed dispersion, however, as the estimated marginal mean decreased in the pretest. Coming to the average distance, the analysis revealed a significant main effect of time, and the estimated marginal mean average distance increased in the posttest, as did distance dispersion. Finally, fuel consumption (based on tractive energy computations; see Appendix 1) also revealed a significant main effect of time. The estimated marginal mean average consumption decreased in the posttest. **Figure 5** shows examples of pretest (above) and posttest (below) speed variations along the circuit for drivers 1 (left) and 3 (right).

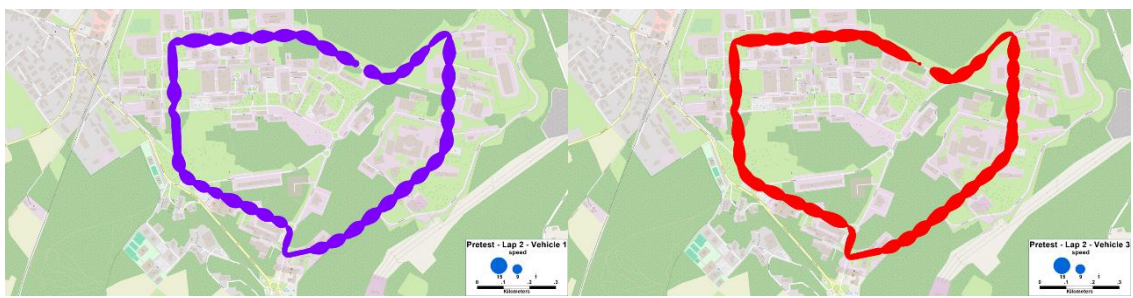

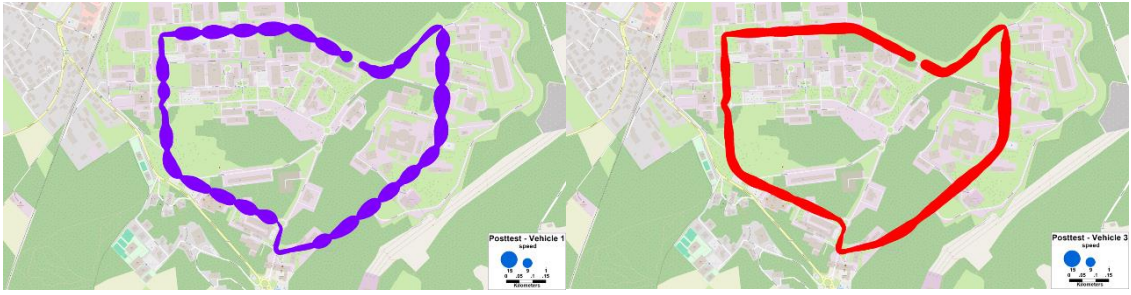

**Fig. 5.** Examples of performances by drivers 1 (blue) and 3 (red) during the pretest (above) and posttest (below)

Finally, a linear mixed-effects model was constructed to examine whether the average road space occupied by the group of drivers changed from the pretest to the posttest (Table 4). These results suggest that the course led to an expansion in inter-vehicle spacing in the posttest. A second linear mixed-effects model was used to assess whether the dispersion of inter-vehicle spacing changed from pretest to posttest. The effect of time was also statistically significant, and the mean dispersion increased, suggesting that, following the course, the spacing between vehicles became not only larger but also more variable. The Platoon results for the circuit are the opposite of those obtained at the WDC (Table 2); this discrepancy will be considered in the discussion.

### 3.3 Safety indicators

This section presents three metrics: average time gap, average time to collision (TTC), and a proactive fuzzy surrogate safety (PFS) metric based on Mattas et al. (2020). PFS identifies tailgating cases in which driving behavior may not be safe but there is no imminent danger. The PFS ranged from 0 to 1, with 0 being totally safe and 1 being totally unsafe. The PFS was calculated according to equation 1:

$$PFS(dist_{lon}) = \begin{cases} 1, & 0 < dist_{lon} - d_1 < d_{unsafe} \\ 0, & dist_{lon} - d_1 > d_{safe} \\ \dfrac{dist_{lon} - d_{safe} - d_1}{d_{unsafe} - d_{safe}}, & d_{unsafe} < dist_{lon} - d_1 < d_{safe} \end{cases} \quad (1)$$

where $dist_{lon}$ is the longitudinal distance between two vehicles, $d_1$ is a safety margin of 2 m, and $d_{safe}$ and $d_{unsafe}$ are calculated according to equations 2-3:

$$d_{safe} = u_{ego,lon}\tau + \frac{u_{ego,lon}^2}{2b_{ego,comf}} - \frac{u_{cut-in,lon}^2}{2b_{cut-in,max}} + d_1 \quad (2)$$

$$d_{unsafe} = u_{ego,lon}\tau + \frac{u_{ego,lon}^2}{2b_{ego,max}} - \frac{u_{cut-in,lon}^2}{2b_{cut-in,max}} \qquad (3)$$

with $u_{ego,lon}$ the velocity of the ego vehicle, $u_{cut-in,lon}$ the velocity of the cutting-in vehicle, $\tau$ the reaction time of the ego vehicle, $b_{ego,comf}$ the comfortable deceleration of the ego vehicle, $b_{ego,max}$ the maximum deceleration of the ego vehicle, $b_{cut-in,max}$ the maximum deceleration of the cutting-in vehicle, and $d_1$ a safety distance margin of 2 m.

The average and standard deviation of the time gap are shown in **Figure 6**. The averages are calculated for drivers 3 to 6 and for all circuit laps.

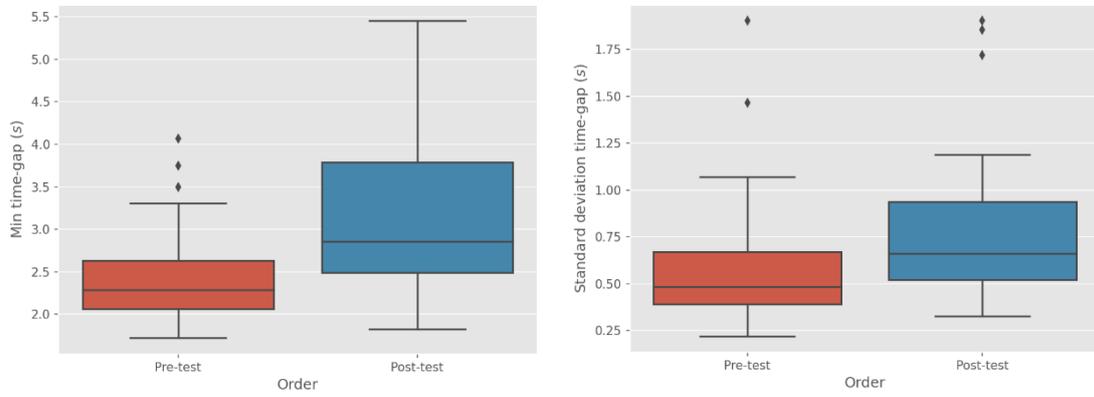

**Fig. 6.** Time gap for all platoon participant drivers and all laps

Both the average and standard deviation of the time gap significantly increase, indicating that on average, the platoon drivers maintain more distance from the vehicle in front of them. **Figure 7 (left)** shows the minimum TTC, averaged for all platoon drivers and all circuit laps, and **Figure 7 (right)** shows the average PFS, again averaged for all platoon drivers and all circuit laps.

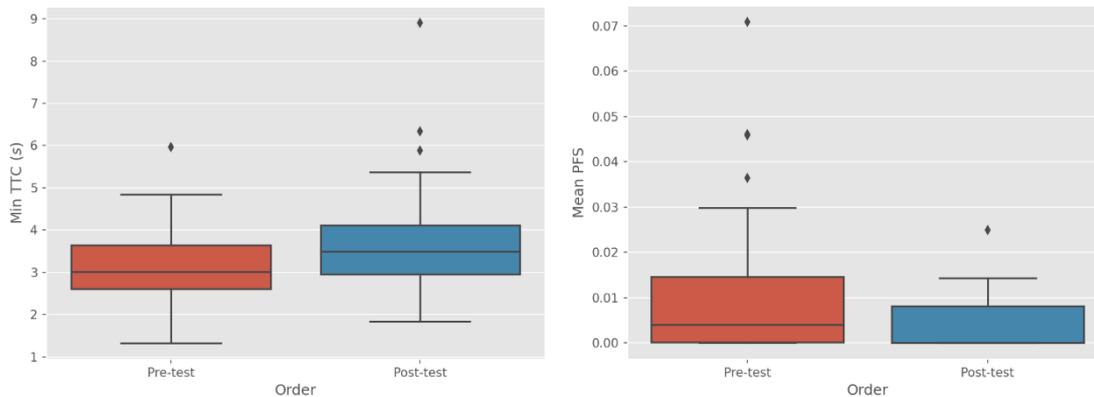

**Fig. 7.** Minimum time to collision (TCC) for all platoon participant drivers and all laps (left) and mean PFS for all platoon participant drivers and all laps (right)

## 4. Discussion

In the pretest phase of this study, the participants (3-6 years old) drove according to the DD principle. This result has been consistently observed in previous studies: the DD technique is the default technique. However, the results also indicate that after the DI course (tutorials + practice), drivers changed their driving patterns in the posttest, both when using the WDC simulator and when driving real cars in the circuit. Our observations from the simulated WDC evaluation indicate that, in the posttest, drivers adopted a DI strategy characterized by reduced speed variability and lower fuel consumption. This, in turn, resulted in a platoon of followers whose inter-vehicle distances fluctuated to a lesser extent. Only the results for the distance following the leader were outside what was expected according to previous studies. Generally, we would expect less car-following distance and less distance dispersion in the pretest and vice versa in the posttest. However, in this case, we observed the absence of differences in the average following distance from the leader and a lower dispersion of the distance in the posttest. After analyzing in detail the ECDs of each participant (Figures 3, 4), we observed some difficulties in adapting the car-following behavior in the pretest. It is possible that the cause of these results is the changes introduced in the handling of the accelerator and brake pedals in the current version of the WDC (designed to also be used on a tablet or smartphone), which is based on the simple position of the mouse over the accelerator and brake on the screen (and not by using the keyboard keys, as usual in previous studies). This topic will undoubtedly receive increased attention in the future.

Focusing on the circuit test, we can see that, as expected after having followed the DI course, driving in the subsequent test returns smoother speed profiles. Although there are some fluctuations in posttest (WDC) speeds, they are much less pronounced than pretest (normal) speeds. These differences also apply to the other drivers in the platoon. The initial conclusion from these results is that the WDC had a positive effect on driving behavior. The results clearly indicate that the absolute acceleration in the posttest experiments is lower than that in the pretest for each driver order. According to both the pretest and the posttest, the standard deviation of the acceleration is greater for drivers 3 and 4 than for drivers 5 and 6. This may be related to the fact that the drivers at the end of the platoon are able to anticipate the behavior of the drivers in front of them. On average, both the absolute and quadratic accelerations decreased after the WD driving course. Similarly, the standard deviations of accelerations also decreased after the WD driving course. This indicates that a smoother driving pattern is observed for platoon drivers but not necessarily a significant decrease in average speed. We observed that both the mean and the standard deviation of the time interval increased significantly, indicating that, on average, the drivers in the platoon maintained a greater distance from the vehicle in front of them. Overall, safety metrics indicated safer driver behavior following WDC. The minimum TTC increases on average from 3.0 to 3.5 seconds, and the PFS approaches zero (indicating completely safe behavior) after learning the DI technique. This evidence is important: although the expected car-following distance patterns were missing during the evaluation in the WDC simulator, the behavior with real cars in the posttest confirmed good transference of the DI technique.

## 5. Conclusions

This study showed the potential for learning DI after a short administration of the WDC course. An empirical exploration of the effectiveness of WDC in driving with real vehicles was presented via a closed-circuit field test. Interestingly, this is the first proven transfer of the effectiveness of WDC in changing behavior with real vehicles and in circumstances similar to driving in heavy traffic. It is also interesting to note that the bases of the course and its graphic design allowed us to cross cultural borders. Until now, the WDC has been tested only in simulations with uniform samples from Israel (Tenenboim et al., 2022) and Spain (Lucas-Alba et al., 2022). However, among the 12 participants in our study, there were drivers from Colombia, Greece, Italy, Hungary, and Spain. The course was taught in English, but its configuration and graphic design are general and totally independent of the spoken language.

It is important to acknowledge that future improvements should be incorporated into this study. Working with a larger sample would be advisable—even though the effect sizes reported in previous studies suggest that the current sample is adequate. In addition, the inclusion of a control group would strengthen the validity of the findings. It would also be beneficial to implement longitudinal designs to assess whether the learning effects persist over time both in the simulator and in real-world driving conditions. We should also improve the technological aspects of data collection and recording. Undoubtedly, the next undisputed improvement would be to test the transfer of WDC to real traffic conditions, which would show the actual potential benefits for our society.

**Acknowledgments**

We thank Paula Fernández, a human resources expert at Alsa, for her help during the development of this pilot study. The authors are also grateful to the twelve volunteers who agreed to participate in the experimental campaign.

**Author contributions**

The authors confirm the contributions to the paper as follows: study conception and design: K. Mattas, T. Toledo, A. Lucas-Alba, O. Melchor, B. Ciuffo; data collection: K. Mattas, A. Lucas-Alba; analysis and interpretation of results: K. Mattas, S. Bekhor, A. Lucas-Alba; draft manuscript preparation: S. Bekhor, A. Lucas-Alba. All the authors reviewed the results and approved the final version of the manuscript.

**Statements and declarations**

Competing interests: The authors declare no competing interests.

**Appendix 1: Computing fuel consumption**

Since we are interested in comparing fuel consumption between pretest and posttest results for the same vehicle, it is sufficient to estimate tractive energy consumption. According to He et al. (2020), the tractive energy consumption (Et, kWh/100 km) of a vehicle is calculated by integrating the tractive power requirements ($P_t$, kW) at the wheels over time, not considering the negative power components from regenerative braking:

$$P_t = \begin{cases} (F_0 + F_1 v_e + F_2 v_e^2 + 1.03 m a_e + mg\sin\theta) v_e 10^{-3}, P_t \geq 0 \\ 0, P_t < 0 \end{cases}$$

$$E_t = \frac{\int_0^T P_t dt}{0.036 \int_0^T v_e dt}$$

where $F_0$, $F_1$ and $F_2$ are road load coefficients; m is the vehicle mass (kg); $v_e$ and $a_e$ are the ego vehicle's speed (m/s) and acceleration (m/s$^2$), respectively; θ is the road gradient (rad); g is the gravitational acceleration (9.81 m/s$^2$); dt is the time interval (s) between consecutive measurement points; and T denotes the total duration (s) of the travel period.

Note that the equation above does not consider the negative power from regenerative braking. In the experiment, all the vehicles had similar fuel characteristics; therefore, the following coefficients were applied: vehicle mass=1360 kg, $F_0$=112.1, $F_1$=0.655, and $F_2$=0.03181. The acceleration is calculated every 0.1 second using the speed data collected in the experiment (He et al., 2020).